\documentclass{emulateapj}

\shorttitle{ENERGY-DEPENDENT VARIABILITY IN BLAZARS}
\shortauthors{Honda}

\begin{document}

\title{PHASE-TRANSIENT HIERARCHICAL TURBULENCE
AS AN ENERGY CORRELATION\\ GENERATOR OF BLAZAR LIGHT CURVES}
\author{Mitsuru~Honda}
\affil{Plasma Astrophysics Laboratory, Institute for Global Science,
Mie 519-5203, Japan}

\begin{abstract}
Hierarchical turbulent structure constituting a jet is considered
to reproduce energy-dependent variability in blazars,
particularly, the correlation between X- and gamma-ray
light curves measured in the TeV blazar Markarian 421.
The scale-invariant filaments are featured by the ordered
magnetic fields that involve hydromagnetic fluctuations
serving as electron scatterers for diffusive shock acceleration,
and the spatial size scales are identified with the local maximum
electron energies, which are reflected in the synchrotron spectral
energy distribution (SED) above the near-infrared/optical break.
The structural transition of filaments is found to be responsible
for the observed change of spectral hysteresis.
\end{abstract}

\keywords{BL Lacertae objects: individual (Mrk 421) ---
galaxies: jets --- magnetic fields ---
radiation mechanisms: nonthermal --- turbulence}

\section{INTRODUCTION}

A noticeable feature associated with blazars is that
the updated shortest variability timescale reaches a few
minutes \citep[e.g., Mrk 421:][]{cui,blazejowski}, not likely to be
reconciled with the light-crossing time at the black hole horizon.
One possible explanation for this fact is that
small-scale structure does exist in the parsec-scale
jet anchored in the galactic core \citep{hh04}.
Indeed, in the plausible circumstance that the successive
impingement of plasma blobs (ejected from the core) into
the jet bulk engenders collisionless shocks, electromagnetic
current filamentation (characterized by the skin depth)
could be prominent \citep{medvedev}.
It is known that the merging of smaller filaments
leads eventually to accumulation of magnetic energy
in larger scales \citep{honda00a,silva}.

Reflecting the self-similar (power law) characteristic
in the inertially cascading range, the local magnetic
intensity of the self-organized filaments will obey
$|{\bf B}|\sim B_{m}(\lambda/d)^{(\beta-1)/2}$,
where $\lambda$ and $d$ reflect the transverse size scale
of a filament and the maximum, respectively,
$B_{m}\equiv |{\bf B}|_{\lambda=d}$, and $\beta$ $(>1)$
corresponds to the filamentary turbulent spectral index.
The value of $d$ is limited by the transverse size of jet
(or blob size; $D$).
Then, it is reasonable to consider that in fluid timescales,
the well-developed coherent fields are sure to actually meet
hydromagnetic disturbance independent of the filamentation;
that is, the turbulent hierarchy is established (see Fig.~1).
The spectral index of the superposed fluctuations
[denoted as $\beta^{\prime}$ $(>1)$] could be different
from $\beta$, and the correlation length scale is
presumably limited by $\sim\lambda$.

At this site, the electrons bound to the local mean fields
suffer scattering by the fluctuations, to be diffusively
accelerated by the collisionless shocks \citep[see][]{hh07}.
When the acceleration and cooling efficiency depend on
the spatial size scales, the local maximum energies of accelerated
electrons will be identified by $\lambda$ (\S\,2), to be reflected
in the synchrotron SED extending to the X-ray region.
More interestingly, the spatially inhomogeneous property
of particle energetics is expected to cause the
energy-dependent variability of broadband SEDs.
Here the naive question arises whether or not this idea is responsible
for the observed elusive patterns of energy correlation of light curves
\citep[e.g.,][]{takahashi,fossatiI,fossatiII,blazejowski}:
this is the original motivation of the current work.

In the present simplistic model, light travel time effects
would still prevent the detection of variability signatures
on timescales shorter than $D/(c\delta_z)$, where
$\delta_{z}=\delta/(1 + z)$, and $\delta$, $z$, and $c$ are
the beaming factor of the jet, redshift, and speed of light, respectively.
However, if a filamented piece is isolated,
having loose causal relation with the dynamics of a
bulk region serving as a dominant emitter, an intrinsic rapid
variability involved in the subsystem would be viable.
Namely, it is inferred that the shorter timescale is
at least potentially realized, and observable, unless
energetic emissions from such a compact domain are crucially
degraded by synchrotron self-absorption and/or $\gamma\gamma$
absorption \citep[e.g.,][]{aharonian}.
As is, the basic notion of the present model seems to provide
a vital clue to settle the debate as to the causality problem
incidental to observed rapid variabilities.

In this Letter, I demonstrate that the hierarchical
system incorporated with the synchrotron self-Compton (SSC)
mechanism accurately generates the time lag of gamma-ray
flaring activity behind the X-ray, confirmed in the
high-frequency-peaked BL\,Lac object Mrk 421 \citep{blazejowski}.
We address that in general, both lag
and lead can appear in X-ray interband correlations,
accompanying the structural transition.
The major transition history is argued in light
of the observed spectral hysteresis patterns.
We also work out $(B_{m},d)$, to provide the constraint
on the field strength and $D$ that should be
compared with those of previous models.

\section{AN IMPROVED EMITTER MODEL WITH\\ HIERARCHICAL STRUCTURE}

We consider a circumstance in which relativistic shocks
propagate through a relativistic jet with the Lorentz
factor $\Gamma$, such that the shock viewed upstream
(jet frame) is weakly to mildly relativistic.
Note the relation of $\delta\sim\Gamma$.
The overall geometry and relative size scales of the
aforementioned hierarchy are sketched in Figure~1.
Provided that the gyrating electrons trapped in the filament
(with the size $\lambda$) are resonantly scattered by
the magnetic fluctuations, the mean acceleration time
upstream is approximately given by
$\tau_{\rm acc}\simeq(3\eta r_{\rm g}/c)[r/(r-1)]$,
where $\eta=(3/2b)(\lambda/2r_{\rm g})^{\beta^{\prime}-1}$,
$b$ is the energy density ratio of fluctuating/local mean
magnetic fields (assumed to be $b\ll 1$),
$r_{\rm g}(\gamma,|{\bf B}|)$ is the electron gyroradius
($\gamma$ being the Lorentz factor), and $r$ is the shock compression ratio.
In the regime in which flares saturate, $\tau_{\rm acc}$
will be comparable to synchrotron cooling time
$\tau_{\rm syn}(\gamma,|{\bf B}|^{2})$.
Balancing these timescales gives the (local) maximum
$\gamma$ of an accelerated electron, described as
$\gamma^{\ast}(\lambda)=\{g_{0}^{-(\beta^{\prime}-1)}g_{1}
\left(\lambda/d\right)^{-[(\beta+1)\beta^{\prime}-2]/2}\}^{1/(3-\beta^{\prime})}$,
where $g_{0}=eB_{m}d/(2m_{e}c^{2})$,
$g_{1}=8\pi^{2}\xi m_{e}^{2}c^{4}/(e^{3}B_{m})$,
$\xi=b(r-1)/r$, and the other notations are standard.
At $\gamma=\gamma^{\ast}$, the electron energy
distribution of the power-law form
$n(\gamma)d\gamma=\kappa\gamma^{-p}d\gamma$ is truncated.

\begin{figure}
\centerline{\includegraphics*[bb=52.0 32.5 382.0 272.0,
width=\columnwidth]{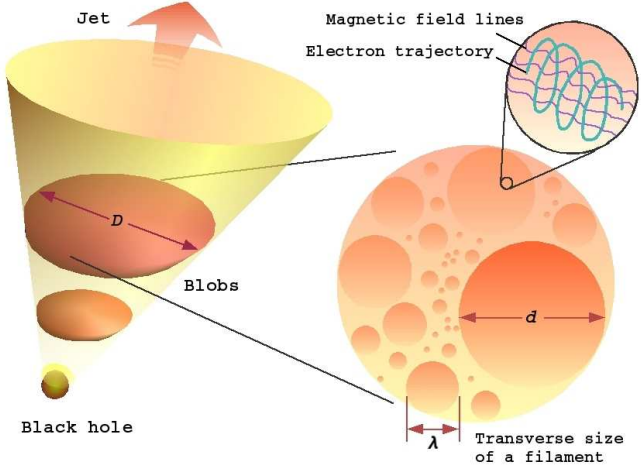}}
\caption{
Schematic of the beamed jet including the emitting blobs and top view
of the transverse cut of a blob region (with the diameter $D$).
A number of circle-like "bubbles" symbolically represent
the transverse section of scale-invariant filaments
(with size $\lambda$, whose maximum is $d$).
Note that $d$ is limited by $D$ (for the values, see \S\,4).
The magnified view of a small sample domain illustrates
the fluctuating magnetic field that scatters
gyrating electrons bound to the local mean field.}
\end{figure}

For simplicity, $\kappa$ and $\xi$ are assumed to be
spatially constant at the moment.
Then, for $\beta^{\prime}<3$ (see \S~3.1), $\gamma^{\ast}$ decreases
as $\lambda$ increases (reflecting likely prolonged $\tau_{\rm acc}$
and shortened $\tau_{\rm syn}$), to take a minimum value at $\lambda=d$,
where the synchrotron flux density makes, up to the frequency of
$(3/4\pi)(\delta_{z}\gamma^{\ast}|_{\lambda=d}^{2}eB_{m}/m_{e}c)(\equiv\nu_{b})$,
a dominant contribution to the $F_{\nu}$ spectrum
(owing to the maximum magnetic intensity at the outer scale).
As $\lambda$ decreases, the flux density tends to decrease,
extending the spectral tail (due to the $\gamma^{\ast}$ increase).
Apparently, this property has the $F_{\nu}$ spectrum steepening
above $\nu_{b}$, whereas below $\nu_{b}$ the spectrum
retains $F_{\nu}\propto\nu^{-(p-1)/2}$.
The frequency $\nu_{b}$ characterizing the
spectral break can be expressed as
$\nu_{b}=7.5\times 10^{14}\delta_{50}
B_{m,10}^{-3/2}\xi_{-4}^{3/2}d_{16}^{-1}~{\rm Hz}$
(for $\beta^{\prime}=5/3$; see \S~3.1), where
$\delta_{50}=\delta_{z}/50$, $B_{m,10}=B_{m}/10\,{\rm G}$,
$d_{16}=d/10^{16}\,{\rm cm}$, $\xi_{-4}=\xi/10^{-4}$,
and $\xi=b(r-1)/r$.

The increase of $\gamma^{\ast}$ in smaller $\lambda$ is
limited at a critical $\lambda_{c}$, below which escape loss
dominates the radiative loss: the equation for the spatial limit,
$r_{g}(\gamma^{\ast})\sim\lambda/2$, yields
$\lambda_{c}/d\sim(g_{0}^{-2}g_{1})^{2/(3\beta+1)}$.
By using this expression, one can evaluate the achievable maximum
$\gamma^{\ast}$ value as
$\gamma^{\ast}|_{\lambda=\lambda_{c}}=g_{0}(\lambda_{c}/d)^{(\beta+1)/2}$,
for which the corresponding synchrotron cutoff frequency is
$\nu_{c}=\delta_{z}\nu_{0}g_{0}^{2}(\lambda_{c}/d)^{(3\beta+1)/2}
=\delta_{z}\nu_{0}g_{1}$,
where $\nu_{0}=(3/4\pi)(eB_{m}/m_{e}c)$.
Also, by combining $\lambda_{c}\propto\nu_{c}^{2/(3\beta+1)}$ with
$F_{\nu_{c}}\propto|{\bf B}(\lambda_{c})|^{(p+1)/2}\nu_{c}^{-(p-1)/2}$,
we read
$\ln(\nu F_{\nu})_{c}/\ln\nu_{c}\propto[(5-p)\beta-(p-1)]/(3\beta+1)$
at $\nu\sim\nu_{c}$.
More speculatively, this scaling might be
reflected in $\ln(\nu F_{\nu})_{p}/\ln\nu_{p}$
for measured synchrotron flux peaks.

\section{PROPERTIES OF ENERGY-DEPENDENT SPECTRAL\\
VARIABILITY AND HYSTERESIS}

\subsection{\it X-Ray Interband Correlation}
In this context, we derive the $\nu$-dependence
of the flaring activity timescale (denoted as $\tau$).
In $\nu_{b}<\nu<\nu_{c}$, which typically covers the X-ray band, we have
$\gamma^{\ast}=[(4\pi/3)(\nu/\delta_{z})(m_{e}c/e|{\bf B}|)]^{1/2}$,
which is written as
$\gamma^{\ast}(\lambda,\nu)=(\nu/\delta_{z}\nu_{0})^{1/2}
(\lambda/d)^{-(\beta-1)/4}$.
Utilizing this, the expression of
$\tau_{\rm syn}(\gamma^{\ast},|{\bf B}|^{2})$ is recast into
$\tau_{\rm syn}(\lambda,\nu)(=\tau_{\rm acc}\sim\delta_{z}\tau)
=(\tau_{0}/\delta_{z})
(\delta_{z}\nu_{0}/\nu)^{1/2}(\lambda/d)^{-3(\beta-1)/4}$,
where $\tau_{0}=36\pi^{2}m_{e}^{3}c^{5}/(e^{4}B_{m}^{2})$.
The relation of $\lambda$ to $\nu$ can be derived from the
equality of $\gamma^{\ast}(\lambda,\nu)=\gamma^{\ast}(\lambda)$,
such that
$\lambda(\nu)/d=\{g_{0}^{-1}g_{1}^{1/(\beta^{\prime}-1)}
(\delta_{z}\nu_{0}/\nu)^{(3-\beta^{\prime})
/[2(\beta^{\prime}-1)]}\}^{4/(3\beta+1)}$.
Substituting this into $\tau(\lambda,\nu)$,
we arrive at the result
$\tau(\nu)\sim(\tau_{0}/\delta_{z})
(\nu/\delta_{z}\nu_{0})^{(\sigma-1)/2}
(g_{0}^{\beta^{\prime}-1}g_{1}^{-1})^{\sigma/(3-\beta^{\prime})}$,
where $\sigma(\beta,\beta^{\prime})
=3(\beta-1)(3-\beta^{\prime})/[(3\beta+1)(\beta^{\prime}-1)]$;
that is,
\begin{equation}
\tau\propto\nu^{-(1/2)(1-\sigma)}.
\label{eq:1}
\end{equation}
Note that $\sigma=0$ (for $\beta=1$) leads to
$\tau\propto\nu^{-1/2}$, formally recovering
the scaling for a homogeneous model.
Significantly, the states of $\sigma<1$ and $>1$ imply the appearance
of the modes for which the X-ray activity in a lower $\nu$ lags
that in a higher $\nu$ \citep["soft lag";][]{takahashi,rebillot}
and vice versa \citep["hard lag";][]{fossatiI}, respectively,
and $\sigma=1$ is the tight-correlation mode \citep{sembay}.
The mode flipping comes about through competing $\lambda$-dependence
of cooling and acceleration efficiency.
In particular, the soft lag appears if
\begin{equation}
\beta<\beta_{c}=(4-\beta^{\prime})/[3(2-\beta^{\prime})].
\label{eq:2}
\end{equation}

\begin{figure}
\centerline{\includegraphics*[bb=70.0 80.0 745.0 460.0,
width=\columnwidth]{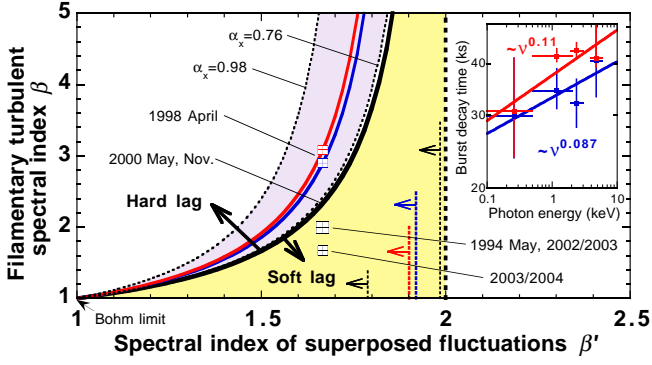}}
\caption{Phase diagram of the hierarchical turbulence.
The function $\beta_{c}(\beta^{\prime})$ (eq.~[\ref{eq:2}];
{\it thick solid curve}) divides the $(\beta,\beta^{\prime})$-plane
into the two domains that allow soft-lag ($\beta<\beta_{c}$;
{\it yellow}) and hard-lag ($\beta>\beta_{c}$) in the X-ray band.
The possible phases $\Phi_{2}$ $(2,5/3)$ and
$\Phi_{5/3}$ $(5/3,5/3)$ ({\it large and small marks,
respectively}) involve soft-lag, as consistent with
the correlations measured in the labeled epochs.
The inset shows the power-law fit to two available
$\tau$-$\nu$ data on 1998 April 21
\citep[for detailed data reduction, see][]{fossatiI},
giving the indices of $s=0.11$ and $0.087$ (corresponding colors),
which determine $\beta(\beta^{\prime};s)$ ({\it colored curves})
for $\beta^{\prime}<(s+2)/(s+1)$ ({\it colored arrows}).
The characteristic curves lie in a domain ({\it purple};
$\beta^{\prime}$-range is indicated by thin solid arrows)
restricted by the measured X-ray spectral indices
\citep[and $p=1.6$;][]{fossatiII}, and give $\beta=3.1$ and $2.9$
({\it colored marks}) at $\beta^{\prime}=5/3$.
For further explanation, see the text.}
\end{figure}

The critical function $\beta_{c}(\beta^{\prime})$ is, for the
key range of $(1<)\beta^{\prime}<2$, plotted in Figure~2.
Note that for the special $\beta^{\prime}=2$ case,
$\sigma[=3(\beta-1)/(3\beta+1)]<1$ is always satisfied,
and $\beta^{\prime}>2$ ensures $\beta>\beta_{c}$
(because of $\beta_{c}<1$): $\beta^{\prime}=2$ and $>2$ lead to soft
and hard lag, respectively, irrespective of the $\beta$ value.
While the index $\beta$ is expected to be variable
(reflecting the long-term structural evolution of
filaments; see \S\,3.3 for details), $\beta^{\prime}$
would be a constant since a mechanism of superimposed magnetic
fluctuations (\S\,1) perhaps has universality.
One can exclude $\beta^{\prime}>2$, which yields
by no means soft lag, which is at odds with the
observational facts, whereupon we can take the modified
upper bound (indicated in Fig.~2, {\it arrows}) into account.
With these ingredients, I conjecture the preferential
appearance of $\beta^{\prime}=5/3$ (the Kolmogorov-type
turbulence), for which $\beta_{c}=7/3$.
As for $\beta$, it appears that for $\beta=2$ \citep{montgomery}
and $5/3$ (and given $\beta^{\prime}=5/3$), the model synchrotron
spectra of $\nu F_{\nu}\propto\nu^{0.32}$ and $\nu^{0.40}$
(in $\nu_{b}<\nu<\nu_{p}$; $\nu F_{\nu}\propto\nu^{0.69}$
in $\nu<\nu_{b}$) provide a reasonable fit to the
measured ones (at flares) in the mid state (2002--2003) and
high state (2004--2005), respectively, of Mrk 421,
suggesting $\nu_{b}\sim 2\times 10^{14}\,{\rm Hz}$
(not shown in figure), as compatible with smaller variability
in the $\nu$-range below {\it R} band \citep{blazejowski}.
Below, we refer to these possible phases
$(\beta,\beta^{\prime})=(2,5/3)$ and $(5/3,5/3)$,
which satisfy equation~(\ref{eq:2}),
as "$\Phi_{2}$" and "$\Phi_{5/3}$", respectively.
Also, we compare with the detailed data of
burst decay time \citep[in 1998;][]{fossatiI}.
The guideline is given in Figure~2: use is made of the
translation of the measured timescale $\propto\nu^{s}$
to $\beta=[4-\beta^{\prime}+s(\beta^{\prime}-1)]/
\{3[2-\beta^{\prime}-s(\beta^{\prime}-1)]\}$.
This characteristic curve for $s\simeq 0.1$ indicates
$\beta\simeq 3$ at $\beta^{\prime}=5/3$, and
$\beta>\beta_{c}$, as consistent with the measured hard lag.
For these $\beta=5/3$, $2$, and $3$, we anticipate
$(\nu F_{\nu})_{p}\propto\nu_{p}^{0.8}$, $\nu_{p}^{0.9}$,
and $\nu_{p}$ \citep[for $p\simeq 1.6$; e.g.,][]{macomb},
amenable to the full X-ray data analysis of
Mrk 421 flares \citep{tramacere}.

Let us now estimate the time lag of a soft energy band
$\epsilon_{\rm L}$($>h\nu_{b}$; $h$ is the Planck constant)
behind a hard band $\epsilon_{\rm H}(<h\nu_{c})$ by
$\Delta\tau=\tau(\epsilon_{\rm L})-\tau(\epsilon_{\rm H})$.
Here it is instructive to note the relation of $(\lambda_{c}<)$
$\lambda(\epsilon_{\rm H})<\lambda(\epsilon_{\rm L})$ $(<d)$.
Using the expression of $\nu_{b}$ (eliminating
$\xi^{3/2}d^{-1}$), we obtain
\begin{equation}
\Delta\tau=1.8\,\delta_{50}^{-1/2}B_{m,10}^{-3/2}
\nu_{b,14}^{-3/7}\epsilon_{{\rm L},1}^{-1/14}\eta_{-1}~{\rm hr}
\label{eq:3}
\end{equation}
\noindent
for the structural phase $\Phi_{2}$, where
$\nu_{b,14}=\nu_{b}/10^{14}\,{\rm Hz}$,
$\eta_{-1}=[1-(\epsilon_{\rm L}/\epsilon_{\rm H})^{1/14}]/10^{-1}$,
and $\epsilon_{{\rm L},1}=\epsilon_{\rm L}/1\,{\rm keV}$.
Concerning the validity, it has been checked that, e.g.,
for a soft-lag episode \citep[in 1994 May;][]{takahashi},
the measured time lag plotted against $\epsilon_{\rm L}$
could be more naturally fitted by the function (\ref{eq:3})
of $\Delta\tau(\epsilon_{\rm L},\epsilon_{\rm H})$
(given $\epsilon_{\rm H}\simeq 4-5\,{\rm keV}$ for {\it ASCA}),
rather than the function of $\sim\epsilon_{\rm L}^{-1/2}
[1-(\epsilon_{\rm L}/\epsilon_{\rm H})^{1/2}]$
for the homogeneous ($\sigma=0$) model.

\subsection{\it X/Gamma-Ray Cross-Band Correlation}
The interband correlation property is reflected in the cross-band
correlation between X- and gamma-rays, provided the SSC mechanism
as a dominant gamma-ray emitter \citep[e.g.,][]{maraschi,dermer}.
Along the heuristic (time independent) manner,
we suppose $\gamma\sim\gamma^{\ast}$ for scattering electrons,
and examine the correlation between an X-ray band $\epsilon_{\rm x}$
(compared to $\epsilon_{\rm H}$) and gamma-ray band
$\epsilon_{\gamma}$ susceptible to the inverse Comptonization
of low-energy synchrotron photons (with $\epsilon_{\rm L}$).
Here we focus on the feasible, Thomson regime of
$(\epsilon_{\rm L}/\delta_{z})\gamma^{\ast}<m_{e}c^{2}$;
note that using the expression of
$\gamma^{\ast}[\lambda(\epsilon_{\rm L})]$ (\S\,3.1),
this range can be written as
$\epsilon_{\rm L}<\delta_{50}[B_{m,10}^{2}(\xi_{-4}^{3/2}
d_{16}^{-1})]^{2/23}~{\rm keV}$ (for $\Phi_{2}$).
The Lorentz factor of the electrons that execute the boost of
$\epsilon_{\gamma}/\epsilon_{\rm L}=(\gamma^{\ast})^{2}$
is denoted as $\gamma_{s}^{\ast}=(\epsilon_{\gamma}/\delta_{z}h\nu_{0})^{1/4}
[\lambda(\epsilon_{\rm L})/d]^{-(\beta-1)/8}$.
Then, simply estimating
$\Delta\tau_{\gamma{\rm x}}
=\tau[\epsilon_{\gamma}/(\gamma_{s}^{\ast})^{2}]-\tau(\epsilon_{\rm x})$
($>0$, for $\beta<\beta_{c}$) would be adequate for the present purpose.
For convenience, one may eliminate $\epsilon_{\rm L}$ from
$\gamma_{s}^{\ast}$ [transform $\lambda(\epsilon_{\rm L})$
into $\lambda(\epsilon_{\gamma})$], and adopt
the positive soft-lag representation of
$\Delta\tau_{{\rm x}\gamma}(=-\Delta\tau_{\gamma{\rm x}})$,
so that the negative sign indicates gamma-ray lag.
Again using $\nu_{b}$, we find for $\Phi_{2}$
\begin{equation}
\Delta\tau_{{\rm x}\gamma}=-1.7\,\delta_{50}^{-5/32}B_{m,10}^{-7/8}
\nu_{b,14}^{-7/16}\epsilon_{\gamma,1}^{-1/32}
\eta_{\gamma{\rm x},-1}~{\rm days},
\label{eq:4}
\end{equation}
\noindent
where $\eta_{\gamma{\rm x},-1}=\{1-0.79\epsilon_{{\rm x},25}^{-1/14}
[(\epsilon_{\gamma,1}\delta_{50}B_{m,10})^{1/2}\times$
$\nu_{b,14}^{1/7}]^{1/16}\}/10^{-1}$,
$\epsilon_{\gamma,1}=\epsilon_{\gamma}/1\,{\rm TeV}$,
and $\epsilon_{{\rm x},25}=\epsilon_{\rm x}/25\,{\rm keV}$.
The simultaneous equations~(\ref{eq:3}) and (\ref{eq:4})
contain the solutions $(\delta,B_{m})$,
for given observable quantities $\nu_{b}$ and
$(\Delta\tau,\Delta\tau_{{\rm x}\gamma})$, as well as
$(\epsilon_{\rm L},\epsilon_{\rm H};\epsilon_{\rm x},\epsilon_{\gamma})$
inherent in detectors.

\begin{figure}
\centerline{\includegraphics*[bb=39.0 40.0 572.0 667.0,
width=\columnwidth]{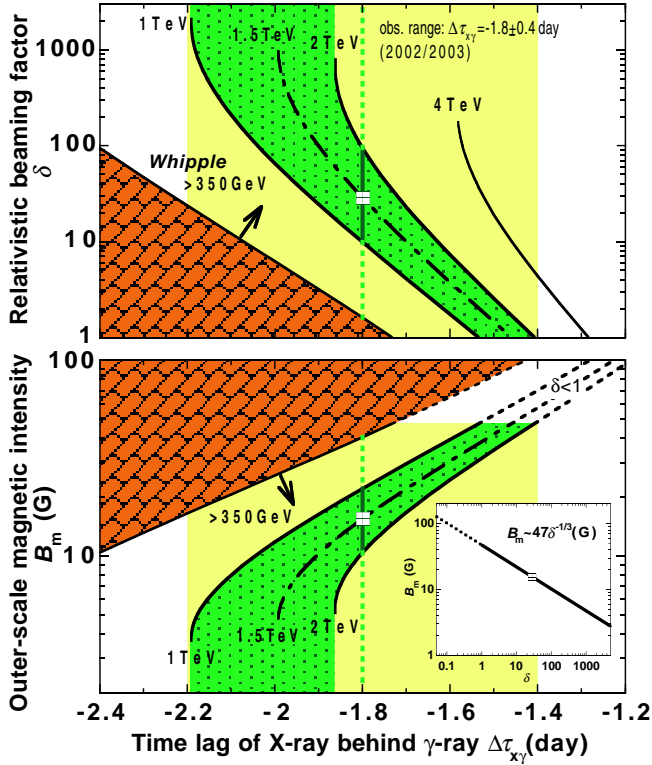}}
\caption{Beaming factor $\delta$ ({\it top}) and maximum field
strength $B_{m}$ ({\it bottom}) vs. the time lag of
X-ray behind gamma-ray $\Delta\tau_{{\rm x}\gamma}$
($<0$ means gamma-ray lag) for a given gamma-ray band
$\epsilon_{\gamma}$ (labeled) as a parameter.
The horizontal axes are common in the panels.
The correlation is taken between $\epsilon_{\rm x}=25\,{\rm keV}$
and $\epsilon_{\gamma}\geq 350\,{\rm GeV}$ compared to an
{\it RXTE} band and coverage of the Whipple\,10\,m telescope
(brick-like shaded area indicates the prohibited domains), respectively.
The hatched bands ({\it green}) indicate the allowed domains
for $\epsilon_{\gamma}=1-2\,{\rm TeV}$ and $\delta>1$, which
cover the measured $\Delta\tau_{{\rm x}\gamma}=-1.8\pm 0.4\,{\rm days}$
({\it yellow}) and give $\delta=10-92$ and $B_{m}=10-22\,{\rm G}$
at $\Delta\tau_{{\rm x}\gamma}=-1.8\,{\rm days}$ ({\it deep green bars};
say, $\delta=29$ and $B_{m}=16\,{\rm G}$ [{\it marks}] for
$\epsilon_{\gamma}=1.5\,{\rm TeV}$ [{\it dot-dashed curves}]).
The inset in the bottom panel shows the $B_{m}$-$\delta$
relation independent of $\epsilon_{\gamma}$.}
\end{figure}

In Figure~3 ({\it top}) for $\nu_{b,14}=2$, $\Delta\tau=1\,{\rm hr}$,
and $(\epsilon_{{\rm L},1},\epsilon_{\rm L}
/\epsilon_{\rm H};\epsilon_{{\rm x},25})=(1,0.2;1)$, compared to
Mrk 421 ($z=0.031$) measurements \citep{takahashi,blazejowski},
the self-consistent numerical solution $\delta$ is plotted against
$\Delta\tau_{{\rm x}\gamma}$, given $\epsilon_{\gamma}$
that covers a gamma-ray band associated with the
Whipple observation \citep[][]{catanese}.
For the allowed domain of $\delta>1$ \citep{piner},
a typical TeV range of $\epsilon_{\gamma,1}\simeq 1-2$
(susceptible to the significant variation in the mid state) is
found to a priori restrict the domain of the observable
$-\Delta\tau_{{\rm x}\gamma}$ to $1.4-2.2\,{\rm days}$.
Surprisingly, this quantitatively agrees with
$\Delta\tau_{{\rm x}\gamma}=-1.8\pm 0.4\,{\rm days}$ that has been revealed
by multiband monitoring in the 2002/2003 season \citep{blazejowski}.
In order to solidify the argument, the solutions for the
high state with $\Phi_{5/3}$ have also been sought.
The results show that the upper bound of $-\Delta\tau_{{\rm x}\gamma}$,
at which $\delta$ diverges, shifts (from $2.2\,{\rm days}$) to
$1.7\,{\rm days}$ and the Whipple coverage $\epsilon_{\gamma,1}<10$
restricts to $-\Delta\tau_{{\rm x}\gamma}>0.7\,{\rm days}$; these
combination yields $-\Delta\tau_{{\rm x}\gamma}\simeq 0.7-1.7\,{\rm days}$.
This is certainly compatible with the measured
$\Delta\tau_{{\rm x}\gamma}=-1.2\pm 0.5\,{\rm days}$
\citep[in the 2003/2004 season; for the significance, see][]{blazejowski}.

\subsection{\it Hysteresis Reversal via Structural Transition}
From the view point of activity history, it is claimed that, involving
the fluctuations with a common $\beta^{\prime}=5/3$, the
coherent structure, at least, in the dominant emission region
has been in the $\beta=2$ phase \citep[1994 May;][]{takahashi},
$\beta=3$ \citep[1998 April;][]{fossatiI,fossatiII}, an intermediate
phase around $\beta=7/3$ \citep[2000 May and November;][]{sembay},
$\beta=2$ \citep[2002/2003 season;][]{blazejowski}, and $\beta=5/3$
\citep[2003/2004 season;][]{blazejowski}, to give rise to a
hard and soft X-ray lag for $\beta\gtrless\beta_{c}=7/3$,
respectively, and no lag for $\beta=\beta_{c}$, as consistent
with the observed correlation properties in each epoch (Fig.~2).
At this juncture, the confirmed reversal between clockwise
\citep{takahashi,rebillot} and anticlockwise \citep{fossatiII}
hysteresis loops in the flux--spectral index plane is ascribed
to the phase transition between $\beta<\beta_{c}$
and $>\beta_{c}$, respectively.
Physically, the likely $\beta=2$ is associated with
the prominence of filamentation \citep{montgomery}.
The smaller $\beta=5/3$ in a high state arguably reflects
strong structural deformation, while the larger $\beta=3$
can be interpreted as the dual-cascade phase of two-dimensional
turbulence (e.g., \citealt{krommes} and references therein)
transverse to pronounced filaments \citep{honda00b}.

\section{DISCUSSION AND CONCLUDING REMARKS}

The practical formula that constrains magnetic field strength
is readily obtained from equation~(\ref{eq:3}), and
in parallel, one for $\Phi_{5/3}$ can be derived as well.
We find the outcome that for $\Phi_{2}$ and $\Phi_{5/3}$,
$B_{m}$ must satisfy
\begin{equation}
B_{m}\delta_{z}^{1/3}=\left\{
\begin{array}{l}
  54\,\nu_{b,14}^{-2/7}
  (\Delta\tau^{-1}\epsilon_{{\rm L},1}^{-1/14}
  \eta_{-1})^{2/3}~{\rm G},\\
  33\,\nu_{b,14}^{-2/9}
  (\Delta\tau^{-1}\epsilon_{{\rm L},1}^{-1/6}
  \eta_{-1}^{\ast})^{2/3}~{\rm G},\\
\end{array}
\right.
\label{eq:5}
\end{equation}
\noindent
respectively, where
$\eta_{-1}^{\ast}=[1-(\epsilon_{\rm L}/\epsilon_{\rm H})^{1/6}]/10^{-1}$
and $\Delta\tau$ is in hours.
In Figure~3 ({\it bottom}), we plot the self-consistent solution
$B_{m}$ (against $\Delta\tau_{{\rm x}\gamma}$; corresponding to
$\delta$-$\Delta\tau_{{\rm x}\gamma}$ in Fig.~3 [{\it top}])
that obeys equation~(\ref{eq:5}) for $\Phi_{2}$ ({\it inset})
with the same parameter values as the top panel.
We see that the observed $\delta>1$
\citep{piner} provides the constraint for which local
magnetic intensity ($|{\bf B}|$) never exceeds
$47\,{\rm G}$ for $\Phi_{2}$ ($51\,{\rm G}$ for $\Phi_{5/3}$).
Whereas a mean magnetic intensity $\bar B$ is not well
defined within the present framework, the obtained
scaling of $B_{m,10}\delta^{1/3}\simeq 5$
seems to be reconciled with the conventional
${\bar B}\delta^{1/3}\simeq 0.1-1\,{\rm G}$
derived from fitting a variety of homogeneous
SSC models to the measured broadband SEDs
\citep[e.g.,][]{ghisellini,tavecchio,krawczynski}.

In turn, the quantity of
$\xi_{-4}^{-3/2}d_{16}=7.5\nu_{b,14}^{-1}
\delta_{50}B_{m,10}^{-3/2}$
(valid for $\beta^{\prime}=5/3$; \S\,2)
is self-consistently determined.
Making use of equation~(\ref{eq:5}) to eliminate $B_{m}$,
we have $d=2.6\times 10^{16}(\delta_{50}\xi_{-4})^{3/2}\,{\rm cm}$
for $\Phi_{2}$
[$2.4\times 10^{16}(\delta_{50}\xi_{-4})^{3/2}\,{\rm cm}$ for $\Phi_{5/3}$],
given the common parameter values (such as $\nu_{b,14}=2$).
To estimate $d$, here we call for another expression,
$\nu_{c}=1.0\times 10^{22}\delta_{50}\xi_{-4}\,{\rm Hz}$
[independent of $(\beta,\beta^{\prime})$; \S\,2].
Using this to eliminate $\delta\xi$ from the $d$-expression,
we obtain the simple scaling of
$d=8.2\times 10^{14}\nu_{c,21}^{3/2}\,{\rm cm}$ for $\Phi_{2}$
($7.5\times 10^{14}\nu_{c,21}^{3/2}\,{\rm cm}$ for $\Phi_{5/3}$),
where $\nu_{c,21}=\nu_{c}/10^{21}\,{\rm Hz}$.
The size $d$ implies the allowable minimum of $D$; e.g.,
$\nu_{c,21}=0.1-10$ (yet involving the large observational
uncertainty) provides $D_{16}\gtrsim 10^{-3}$ to $1$
(where $D_{16}=D/10^{16}\,{\rm cm}$),
as reconciled with the previous results
\citep[e.g.,][]{fossatiII,krawczynski,blazejowski}.
It also turns out, from the $\nu_{c}$-scaling, that
the range of $\nu_{c,21}<10^{2}$ accommodates $\xi\ll 1$,
and thereby the assumption of $b\ll 1$ (\S\,2).

In addition, given an energy input into the jet,
particle density $n$ is estimated.
Assuming that electron injection operates at
$\gamma_{\rm inj}\ll\gamma^{\ast}|_{\lambda=d}$ $(\leq\gamma^{\ast})$,
we approximately get $n\simeq(\kappa/0.6)\gamma_{\rm inj}^{-0.6}$
(for $p=1.6$), to find that the steady luminosity
of $10^{44}\,{\rm ergs\,s^{-1}}$, which appears to retain
a dominant portion around the $\nu_{b}$, requires
$n\gtrsim 6\times 10^{4}\gamma_{\rm inj}^{-0.6}
D_{16}^{-3}B_{m,10}^{-1.3}\,{\rm cm^{-3}}$
(when supposing a spherical emitting
volume with the diameter of $D$).
Recalling $B_{m,10}\lesssim 5$, we thus read
$n\gtrsim 10^{3}D_{16}^{-3}\,{\rm cm^{-3}}$
for ordinary $\gamma_{\rm inj}\sim O(1)$; note that
an upper bound can be given by imposing the conditions of,
e.g., pair-plasma production ($T\gtrsim 1\,{\rm MeV}$)
and radial confinement ($nT\lesssim B_{m}^{2}/8\pi$),
such that $n\lesssim 10^{8}\,{\rm cm^{-3}}$
(suggesting $D_{16}\gtrsim 10^{-2}$).

In conclusion, the gamma-ray lags of $1-2\,{\rm days}$
measured in Mrk 421 have been nicely reproduced
by the hierarchical turbulent model of a jet.
The crucial finding is that the structural transition
$\Phi_{2}\rightarrow\Phi_{5/3}$ results in downshifting
the upper bound of the observable lag
[in a TeV ($\epsilon_{\gamma,1}\simeq 1$) band]
from $2.2$ to $1.7\,{\rm days}$, in accordance with a
closer inspection from 2002 to 2004 by \citet{blazejowski}.
A typical $1.8\,{\rm day}$ lag (in the 2002/2003 season) suggests
$\delta=10-92$ and $B_{m}=10-22\,{\rm G}$ (Fig.~3);
the latter provides an upper limit of local magnetic intensity.
The present model as a possible alternative to the previous
leptonic \citep[e.g.,][]{sikora,bednarek,konopelko} and
hadronic scenarios \citep[e.g.,][]{muecke} will shed light
on puzzling aspects of broadband spectral variability.

\end{document}